\documentclass[aps,pre,twocolumn,groupedaddress]{revtex4-1}

\usepackage{graphicx}
\usepackage{color}
\usepackage{amsmath}

\def\be{\begin{equation}}
\def\en{\end{equation}}
\def\bea{\begin{eqnarray}}
\def\ena{\end{eqnarray}}

\def\ve{\varepsilon}
\def\n{\nabla}

\newcommand{\av}[1]{\langle{#1}\rangle}

\newcommand{\bi}[1]{\mbox{\boldmath$#1$}}
\newcommand{\pp}[2]{\frac{\partial {#1}}{\partial {#2}}}

\begin{document}

\title{
Dynamic coupling between a multistable defect pattern and 
flow in nematic liquid crystals confined in a porous medium 
}

\author{Takeaki Araki}
\affiliation{Department of Physics, Kyoto University, Kyoto 606-8502, Japan}


\begin{abstract}

When a nematic liquid crystal is confined in a porous medium 
with strong anchoring conditions, topological defects, called disclinations, 
are stably formed 
with numerous possible configurations. 
Since the energy barriers between them are large enough, 
the system shows multistability. 
Our lattice Boltzmann simulations demonstrate 
dynamic couplings between 
the multistable defect pattern and the flow in a 
regular porous matrix. 
At sufficiently low flow speed, 
the topological defects are pinned at the quiescent positions. 
As the flow speed is increased, 
the defects show cyclic motions and nonlinear rheological properties, 
which depend on 
whether or not they are topologically constrained 
in the porous networks. 
In addition, we discovered that 
the defect pattern can be controlled by controlling the flow. 
Thus, the flow path is recorded in the porous channels 
owing to the multistability of the defect patterns.

\end{abstract}

\maketitle

Topological defects are observed in various types 
of phases associated with 
spontaneous symmetry breaking such as 
crystals and ferromagnetic materials 
\cite{Chaikin_book_1995}. 
In the early stage of the phase transitions,  
many defects are created, and 
their total amount then decreases with time to 
reduce the free energy \cite{Onuki_book_2002}. 
If frustrations are internally included 
or imposed by external fields, 
the defects are sometimes stabilized. 
On the other hand, re-organization of topological defects 
is often important 
under non-equilibrium conditions. 
For instance, pinning and depinning of quantized 
vortices in a type II-superconductor induce 
electrical resistance \cite{Fisher_PRB_1991}. 
Motions of quantized vortices also lead to mutual 
frictional interactions in a flowing 
superfluid \cite{Onuki_book_2002}.

A nematic liquid crystal (NLC) 
is an ideal system for studying behaviors of topological defects 
because of its slow dynamics and large lengthscale 
\cite{Chuang_Sci_1991,Crawford_book_1996,Musevic_NM_2011}. 
When NLCs are confined in cavities
\cite{Bradac_PRE_1998,Serra_SM_2011}, 
microfluidic devices \cite{Sengupta_SM_2011,Sengupta_MN_2012} 
and porous media 
\cite{Iannacchione_PRL_1993,
Wu_PRL_1992,
Bellini_PRE_1996,Bellini_PRL_2000,Bellini_PRL_2002,Buscaglia_PRE_2006,
Gruener_JPCM_2011}, 
the anchoring interaction between the director 
and the solid surface imposes frustrations on the 
elastic field. 
In a complex geometry, 
many topological defects in NLCs are thus sustained 
for a long time 
\cite{Musevic_SM_2008,Musevic_NM_2011,
Rotunno_PRL_2005,Araki_PRL_2006,Serra_SM_2011,Araki_NM_2011}. 
A NLC in a porous material exhibits 
slow glassy behavior and the resulting memory effects 
\cite{Bellini_PRL_2002,Rotunno_PRL_2005,Buscaglia_PRE_2006}. 
In a previous paper, 
we studied the memory effect by Monte Carlo simulations, 
focusing on the role of the bicontinuity of the matrix \cite{Araki_NM_2011}. 
After cooling from an isotropic state, 
many disclination lines running 
through the channels of the porous network remain, 
with numerous possible trajectories. 
Because each trajectory is at one of the local energy 
minimum states and the energy barriers among them 
are often larger than the thermal energy, 
the system shows non-equilibrium behaviors similar to those in spin glasses. 
We found that the memorization is attributed to 
the re-configurations of the defect structure. 
The applications of an electric or magnetic field 
can bring the system into a new different state, 
and the new configuration remains even after the 
field is removed. 
Here, irreducible disclinations, which are 
topologically concatenated to the solid matrix, 
tend to enhance the memory effect.

In this Letter, we study a NLC 
flowing in a regular porous medium. 
In particular, we focus on 
dynamic couplings between the flow and the nonergodic 
disclination pattern. 
We hope our findings will facilitate research
on microfluidics using liquid crystals 
\cite{Sengupta_SM_2011,Sengupta_MN_2012}.

First, we prepare a pattern for a porous material by 
introducing a scalar variable $\phi$ in 
a cubic lattice according to Ref.~\cite{Araki_NM_2011}. 
$\phi$ varies smoothly in space. 
Then, we partition the cubic lattice into three portions by 
using $\phi$: 
$\mathcal{P}$ is the ensemble of lattice sites representing 
the solid material of the porous matrix; 
$\mathcal{N}$ denotes the nematic fraction 
and $\mathcal{S}$ is the ensemble of lattice sites 
at the interface. 
For the portions $\mathcal{N}$ and $\mathcal{S}$, 
we introduce a nematic tensorial order parameter $Q_{\alpha\beta}$ 
\cite{deGennes_book_1993}. 
We employ the Landau$-$de Gennes free energy density \cite{Qian_PRE_1998},
\bea
&&f_{\rm LdG}(Q_{\alpha\beta})
=\frac{\alpha_{\rm F}}{2}  Q_{\alpha\beta}Q_{\beta\alpha}
-\beta_{\rm F}Q_{\alpha\beta}Q_{\beta\mu}Q_{\mu\alpha}
\nonumber\\
&&
+\gamma_{\rm F}(Q_{\alpha\beta}Q_{\beta\alpha})^2
+\frac{L}{2}(\n_\mu Q_{\alpha\beta})^2
-\Delta \ve 
E_\alpha E_\beta Q_{\alpha\beta}, 
\label{eq:Landau_deGennes}
\ena
where $\alpha_{\rm F}$, $\beta_{\rm F}$, 
$\gamma_{\rm F}$, $L$ and $\Delta \ve$ are 
phenomenological material constants. 
$\bi{E}$ is an external field. 
The repeated subscripts indicate 
summations over Cartesian indices. 
The free energy functional of the system is given by 
\bea
\mathcal{F}\{Q_{\alpha\beta}\}
=\int_{\mathcal{N}+\mathcal{S}}d\bi{r}f_{\rm LdG}(Q_{\alpha\beta})
-w\int_{\mathcal{S}}d\bi{r} Q_{\alpha\beta}d_\alpha d_\beta,  
\ena
where $\bi{d}$ is the normal vector of the porous matrix, 
$\bi{d}=\n\phi/|\n\phi|$.  
The second term represents the molecular anchoring interaction 
between the surface of the porous matrix and the nematic 
orientational order, and $w$ is its material constant. 
We assume $w>0$, 
{\it i.e.}, that homeotropic anchoring is preferred.

The hydrodynamic flow $\bi{u}$ is determined by a momentum 
equation \cite{Qian_PRE_1998}, 
\bea
\rho \left(\pp{}{t}+u_\mu\n_\mu \right)u_\alpha
=\nabla_\beta (-p \delta_{\alpha\beta}+\sigma^{\rm d}_{\alpha\beta}
+\sigma'_{\alpha\beta}), 
\ena
where $\rho$ is the material density and $p$ is the pressure. 
Here, $\sigma^{\rm d}_{\alpha\beta}$ is the distortion stress tensor, 
which is given by 
\bea
\sigma^{\rm d}_{\alpha\beta}=-
\frac{\partial f_{\rm LdG}}{\partial (\nabla_\alpha Q_{\mu\nu})}
(\nabla_\beta Q_{\mu\nu}).
\ena
Further, 
$\sigma'_{\alpha\beta}$ is the viscous stress tensor, which is given by 
\bea
\sigma'_{\alpha\beta}
=&&\beta_1 Q_{\alpha\beta}Q_{\mu\nu}A_{\mu\nu}
+\beta_4 A_{\alpha\beta}
+\beta_5 Q_{\alpha\mu}A_{\mu\beta}
+\beta_6 Q_{\beta\mu}A_{\mu\alpha}\nonumber\\
&&+\frac{1}{2}\mu_2 N_{\alpha\beta}
-\mu_1 Q_{\alpha\mu}N_{\mu\beta}
+\mu_1 Q_{\beta\mu}N_{\mu\alpha},
\label{eq:vis_stress}
\ena
where $A_{\alpha\beta}=(\nabla_\alpha u_\beta+\nabla_\beta u_\alpha)/2$ 
is the symmetric velocity gradient tensor, 
and $N_{\alpha\beta}$ is the time rate of change of $Q_{\alpha\beta}$ 
with respect to the background flow. 
The latter is defined as
\bea
N_{\alpha\beta}=\pp{}{t}Q_{\alpha\beta}
+u_\mu\nabla_\mu Q_{\alpha \beta}
+\omega_{\alpha\mu}Q_{\mu\beta}-Q_{\alpha\mu}\omega_{\mu\beta}, 
\ena
where $\omega_{\alpha\beta}
=(\nabla_\alpha u_\beta-\nabla_\beta u_\alpha)/2$ 
is the asymmetric velocity gradient. 
In Eq.~(\ref{eq:vis_stress}), 
$\beta_i$ ($i=1,4,5,6)$ and $\mu_i$ ($i=1,2$) are 
phenomenological parameters having the dimension of viscosity. 
The time development of the nematic order parameter is 
given by 
\bea
N_{\alpha\beta}
=
-\frac{1}{\mu_1}\frac{\delta\mathcal{F}}
{\delta Q_{\alpha\beta}}-\frac{\mu_2}{2\mu_1}A_{\alpha\beta}. 
\ena
The above nemato-hydrodynamic equations are solved 
with non-slip boundary conditions at the surfaces of the porous matrix.

Lattice Boltzmann simulation is one of the 
most efficient numerical methods 
of investigating flow behaviors in a complex geometry \cite{Succi_book_2001}. 
The bounce-back method realizes non-slip boundary conditions 
on the velocity fields at the surfaces of the porous matrix 
\cite{Succi_book_2001}. 
To our knowledge, two lattice Boltzmann algorithms 
for simulating nemato-hydrodynamics 
have been proposed 
\cite{
Denniston_PRE_2001a,Care_JPCM_2000,Care_PRE_2003,
Spencer_PRE_2006}. 
In this study, we adopt a numerical scheme developed by Care and coworkers. 
Here, we omit the details and follow their notations 
(see Refs.~\cite{Care_JPCM_2000,Care_PRE_2003}). 
We performed three-dimensional simulations 
with four speeds and coordination number 27 (D3Q27). 
The weight factors of the four speeds $t_i$ are given by 
$t_i=8/27$ for $\bi{c}_i=(0,0,0)$, 
$t_i=2/27$ for $\bi{c}_i=(\pm 1,0,0)$, 
$t_i=1/54$ for $\bi{c}_i=(\pm 1,\pm 1,0)$ 
and $t_i=1/216$ for $\bi{c}_i=(\pm 1,\pm 1,\pm 1)$ 
with appropriate permutations. 
Here, $\bi{c}_i$ is the non-dimensional velocity vector. 

In our simulations, we impose an external force 
density $\bi{f}$ to flow the NLC. 
Then, we modify Eq.~(31) in Ref.~\cite{Care_JPCM_2000} to 
\bea
&&\sum_{j}M_{i\alpha\beta j \mu\nu}g_{j\mu\nu}^{\rm (neq)}
=-\frac{g_{i\alpha\beta}^{\rm (neq)}}{\tau_i}+\delta m_{i\alpha\beta}
+\delta p_{i \alpha\beta}\nonumber\\
&&+\sum_{i}\rho t_i\{f_\mu(c_{i\mu}-u_\mu)
+ f_\mu u_\nu c_{i\mu}c_{i\nu}\}\delta_{\alpha\beta}, 
\ena
where $\sum_j$ represents summation over 27 velocity vectors $\bi{c}_j$. 
We employ 
$\alpha_{\rm F}=-0.133$, 
$\beta_{\rm F}=2.133$, 
$\gamma_{\rm F}=1.6$, 
$w=0.1$, 
$L=0.1$, 
$\Delta\ve=1.0$, 
$\mu_1=3.0$, 
$\mu_2=-1.0$, 
$\tau_0=1.1$, 
$\eta_0=0.3$, and 
$\eta_2=0.5$. 
This set of parameters gives 
a correlation length $\xi=0.87\Delta$, 
an anchoring penetration length $d_{\rm a}=\Delta$, 
and 
a bulk nematic order $Q_{\rm b}=0.37$, 
where $\Delta$ is the unit cell length. 
The resulting viscosity ratio of this anisotropic liquid is 
$\eta_\perp/\eta_\parallel\cong 1.1$. 
Here $\eta_\parallel$ and $\eta_\perp$ are the effective viscosities 
when the director is parallel to the flow 
and parallel to the velocity gradient, respectively. 
We adopt a bicontinuous pattern with cubic 
symmetry, which we have studied and is denoted as BC 
in Ref.~\cite{Araki_NM_2011}. 
The size of the unit cell of the bicontinuous cubic is $L=32\Delta$. 
We use a simulation box of $64^3$ 
with periodic boundary conditions. 

In NLCs, the flow speed is characterized well by the Ericksen number $Er$, 
which is the ratio of the viscous and the elastic forces. 
It is given by $Er=|\bi{f}|\ell^3/K$, where $K$ is the Frank elastic 
modulus expressed as $K=9LQ_{\rm b}^2/2$ in a one-constant approximation, 
and $\ell$ is the characteristic length of the distortion. 
Because it is 
difficult to determine $\ell$ quantitatively in a complex geometry, 
we assume that $\ell$ is of the order of the radius 
of the narrowest channels, namely, $\ell=L/4$.  

\begin{figure}[htbp] 
\includegraphics[width=0.35\textwidth]{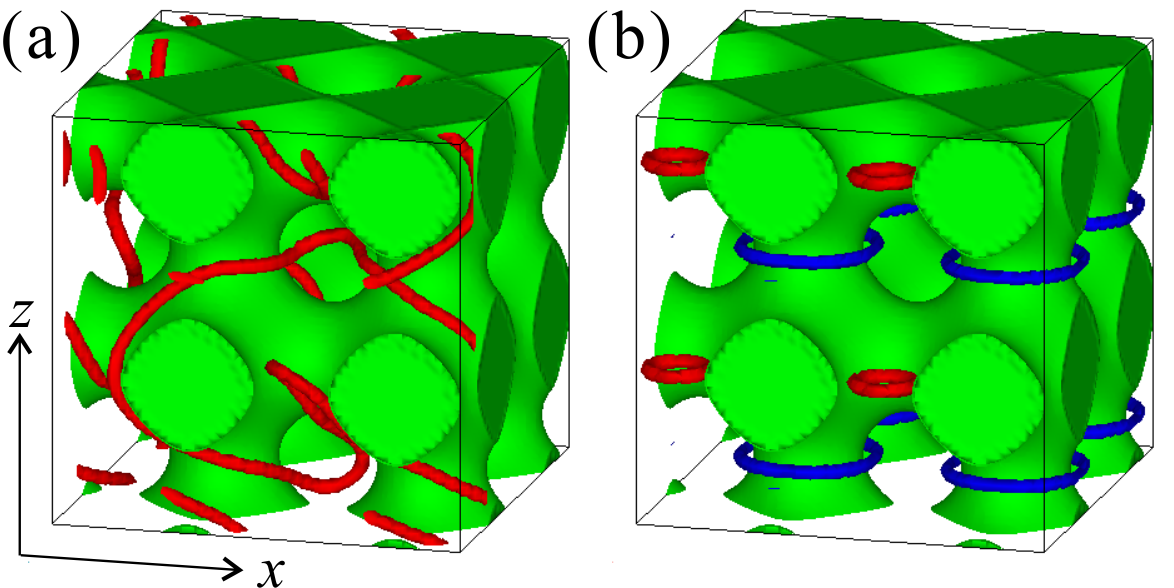}
\caption{
(a) Snapshot of disclination lines of nematic liquid crystals 
in a porous medium after zero field cooling from an isotropic state. 
Solid green objects represent the porous matrix, and 
red curves are the remaining disclinations. 
(b) Defect pattern remaining after an external field pulse is applied
along the $z$-axis. 
Red and blue curves represent 
reducible (type-I) and irreducible (type-II) disclinations, respectively. 
}
\label{fig1}
\end{figure}

We quenched the system from an isotropic state to a nematic 
state without any external field $\bi{E}$ or bulk force $\bi{f}$. 
Because of the anchoring effect, the director was strongly 
deformed, therefore, disclination lines remained 
even after long duration annealing. 
Figure~\ref{fig1}(a) shows one of the stable defect configurations, 
where the defect positions 
are represented by the isosurfaces of the elastic energy density 
$f_{\rm el}=L(\n_\mu  Q_{\alpha\beta})^2/2$. 
Here, the remaining disclinations run through the channels rather randomly. 
They are closed without ends, and some are concatenated 
to the solid matrix. 
Then, we applied a strong external field pulse 
along the $z$-axis ($E_z=0.1$). 
After the field was removed, 
two types of disclination loops formed 
in the $x$-$y$ planes, as shown in Fig.~\ref{fig1}(b). 
Half of the remaining disclinations (red loops in Fig.~\ref{fig1}(b)) 
are localized at the 
narrowest sections of the channels. 
The line defects locally have the topological charge $s=-1/2$, 
so the loops are reducible continuously 
into hyperbolic point defects with 
$s=-1$. 
On the other hand, 
the other half of the loops (blue loops in Fig.~\ref{fig1}(b)) 
consists of disclinations, which locally have the topological 
charge $s=1/2$. 
However, they are irreducible to point defects, 
because the solid necks of the BC penetrate the loops. 
Hereafter, the former and latter 
defect groups are referred to 
as type-I and II, respectively. 
This structure is the same as that obtained 
in the Monte Carlo simulations \cite{Araki_NM_2011}. 
It also appears 
that this configuration is probably 
one of the global energy minimum states, 
which are also degenerated along the $x$ and $y$-axes. 
We employ this configuration as an initial state.  

\begin{figure}[htbp] 
\includegraphics[width=0.35\textwidth]{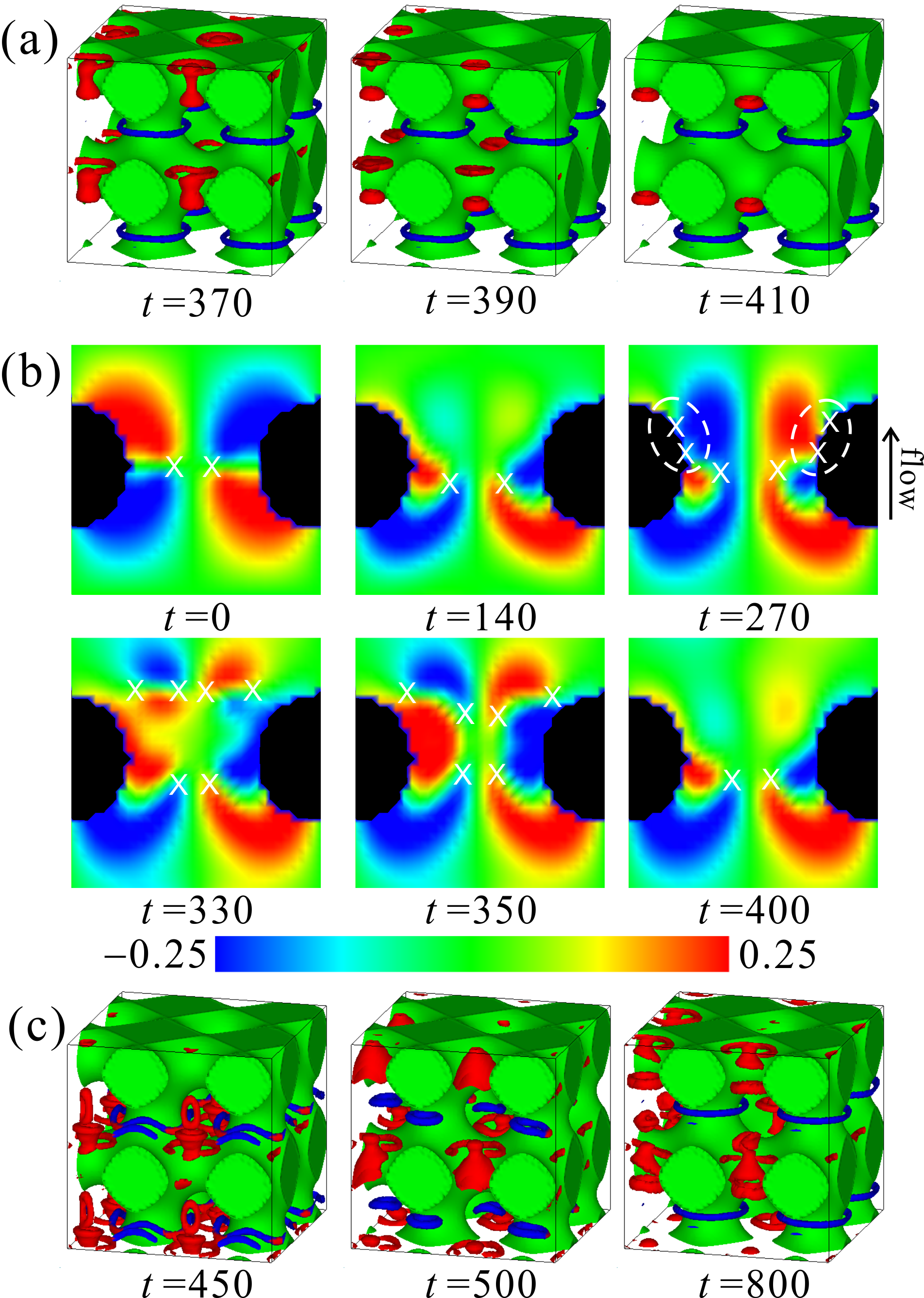}
\caption{
Time evolutions of 
(a) the defect pattern and 
(b) $Q_{zx}$ at an $x$-$z$ plane 
under $f_z=0.006$. 
In (b), the black areas are the solid matrix, 
and white crosses indicate the defect positions. 
At $t=270$, the anchoring condition is broken.
(c) Time evolution of the defect pattern under 
$f_z=0.013$. 
The type-II (blue) defects also exhibit cyclic behavior. 
}
\label{fig2}
\end{figure}

First, we impose force $f_z$ to flow the NLC toward 
the $z$-direction. 
In NLCs, defect motion is caused by not only hydrodynamic 
convection, but also rotations of the director field. 
If the imposed force is weak enough, 
both types of defects show only small displacements, 
which are determined by the balance between 
hydrodynamic convection and director rotation 
(not shown here). 
As the flow speed is increased, 
the director rotation can no longer compensate for 
the hydrodynamic convection, so 
the defect pattern becomes unstable against the flow. 
Figure~\ref{fig2} shows 
the temporal changes in (a) the defect pattern and 
(b) the cross section of $Q_{zx}$ at an $x-z$ plane 
that cuts the centers of the necks of the BC. 
The applied force strength is $f_z=0.006$, which corresponds 
to $Er\cong 50$. 
Under this force, 
the type-I defects start to exhibit cyclic behavior. 
As shown in Fig.~\ref{fig2}(b), 
the director field is gradually distorted with time. 
When the elastic distortion energy becomes comparable 
to the anchoring energy, 
the anchoring at the downstream side of the porous surface 
is transiently broken (see $t=270$ in Fig.~\ref{fig2}(b)). 
When the anchoring breaks, new defects are born from the surface. 
Then, the new defects and the old type-I loops fuse 
into new type-I disclination loops. 
During this cycle, the type-II loops keep to be at the quiescent 
positions. 
With a further increase in the flow speed, 
the type-II loops also become unstable. 
Figure~\ref{fig2}(c) shows snapshots of the defect pattern 
at $f_z=0.013$. 
Because they are concatenated to the channels, 
they cannot move without a topological transformation. 
The type-II defects repeat another cycle, in which 
they are deformed, scissored and reconnected sequentially.

\begin{figure}[htbp] 
\includegraphics[width=0.45\textwidth]{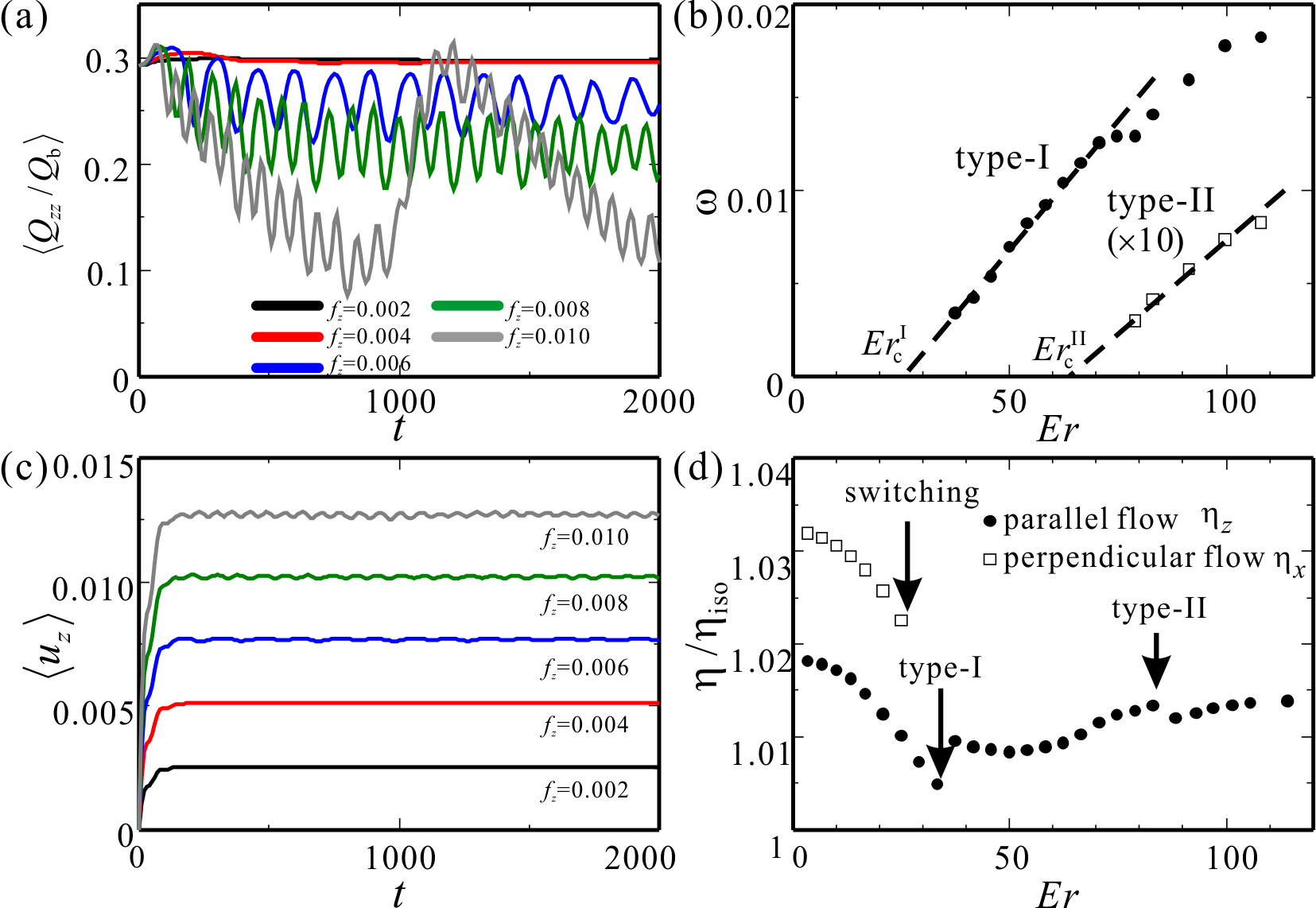}
\caption{
(a) Temporal changes in 
the averaged remnant order $\av{Q_{zz}}$. 
(b) Frequencies of the cyclic motions 
of the type-I and II defects 
as a function of the 
Ericksen number. 
Broken lines are 
provided for better visualization.
(c) Temporal changes in the flow speed 
$\av{u_z}$ of the liquid crystal 
in the porous matrix. 
The force density is increased from $f_z=0.002$ to $0.01$. 
(d) Effective viscosities for the 
parallel and perpendicular flows 
plotted with respect to $Er$. 
The viscosity is normalized by that of an 
isotropic liquid $\eta_{\rm iso}$. 
At the arrows, the defect patterns are re-organized. 
}
\label{fig3}
\end{figure}

In Fig.~\ref{fig3}(a), we plot the temporal changes 
in the averaged orientational order along the $z$-axis $\av{Q_{zz}}$. 
When the force is weak ($f_z\lesssim 0.004$), 
the orientational order is almost fixed at 
the same value as that at rest. 
Subject to an intermediate force ($0.006 \lesssim f_z\lesssim 0.008$), 
the orientational order shows an oscillation mode, 
which corresponds to 
the repeated cycle of the type-I defects  shown in 
Figs.~\ref{fig2}(a) and (b). 
For strong force ($f_z\gtrsim 0.010$), 
two oscillation modes are observed in $\av{Q_{zz}}$. 
The mode of the shorter period 
is the same as that for $f_z=0.006$; 
on the other hand, 
that of the longer period 
corresponds to the cycle for the type-II defects 
shown in Fig.~\ref{fig2}(c). 
The frequencies of the two oscillation modes 
are shown in Fig.~\ref{fig3}(b). 
Both frequencies decreased 
with decreasing $Er$. 
It is likely that they show critical behaviors 
as $\omega\propto |Er-Er_{\rm c}|^k$ with $k=1.0$. 
The critical values of $Er$ are estimated 
to be $Er_{\rm c}^{\rm I}\cong 26$ for the type-I defects and 
$Er_{\rm c}^{\rm II}\cong 64$ for type-II defects by 
extrapolating straight lines. 
However, 
the true critical points are not accessible 
because of our simulation's constraints. 
Therefore, we cannot yet conclude whether 
these bifurcations are continuous or not, 
and determine the exponent $k$.

Figure~\ref{fig3}(c) 
shows the temporal changes in 
the flow speeds $\av{u_z}$ averaged inside the porous matrix. 
Above the critical strength for the cyclic defect behavior, 
the flow speed exhibits oscillation, 
the period of which coincides with that of $\av{Q_{zz}}$. 
In Fig.~\ref{fig3}(d), we plot the apparent viscosity 
$\eta_z$, which is obtained by 
averaging it over $1000\le t\le 2000$. 
Here, the viscosity is normalized by that of an isotropic 
liquid $\eta_{\rm iso}$, which 
is determined by the same simulation method. 
Below the critical force, 
the apparent viscosity 
decreases marginally with $Er$. 
That is, the NLC exhibits a shear-thinning rheology. 
At the two critical values of $Er$ for the onset of the oscillations, 
small kinks in the apparent viscosity are observed, 
indicating that the defect pattern influences the flow properties.

\begin{figure}[htbp] 
\includegraphics[width=0.35\textwidth]{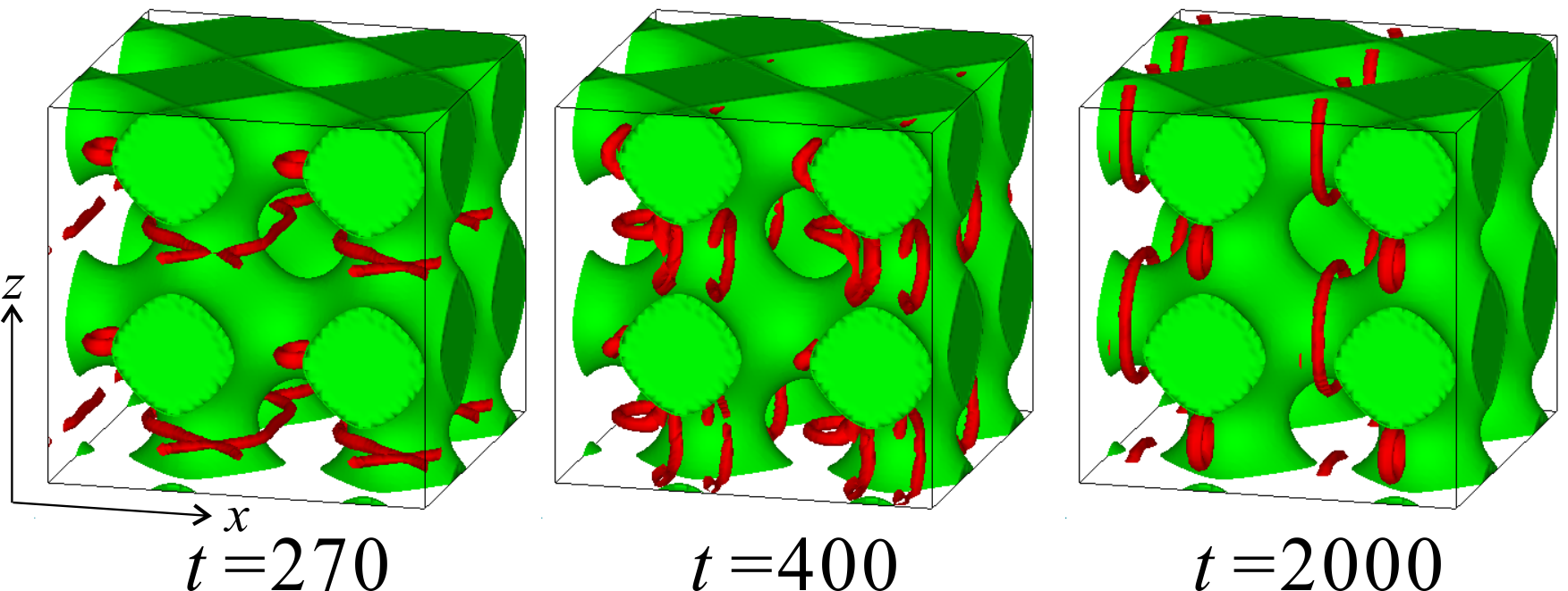}
\caption{
Switching behaviors of topological defects under a 
perpendicular flow. 
A force $f_x=0.005$ is imposed for the defect pattern 
aligned along the $z$-axis. 
}
\label{fig4}
\end{figure}

Next, we impose an external force toward the $x$-direction. 
Under a weak force, 
all the defects are pinned at their original positions, 
showing marginal elongations along the flow (not shown here). 
Above a certain force, 
we found that the defect pattern 
switched to an orientation along the flow direction. 
The switching process of the defect pattern 
under $f_x=0.005$, which corresponds to $Er\cong 42$, 
is shown in Fig.~\ref{fig4}. 
Once the orientation is switched, 
the new defect pattern is sustained 
owing to its multistability 
even after the flow is stopped. 
The apparent viscosity for the $x$-oriented flow, $\eta_x$, 
is also plotted in Fig.~\ref{fig3}(d). 
Below the critical Ericksen number for switching, 
the apparent viscosity along the $x$-axis is larger than 
$\eta_z$. 
At the critical Ericksen number ($Er_{\rm c\perp}\cong 25)$, 
$\eta_x$ decreases abruptly to $\eta_z$. 
After switching, 
the system exhibits 
the same oscillations as those for the parallel flow 
in Fig.~\ref{fig2}. 

Note that 
the oscillation frequencies and the corresponding Ericksen number 
depend on the anchoring strength, 
because the anchoring breaking plays a crucial role 
in the cyclic behaviors of the topological defects. 
Thus, the values displayed are not universal for any anchoring 
strength and pore size. 
Moreover, qualitatively distinct behaviors may be observed 
in extreme cases. 
For instance, defects may be transported unboundedly without the 
cyclic motions if the anchoring strength is infinitely large, 
although our simulations cannot access this regime. 
Because the Ericksen number does not factor in the anchoring effect, 
it cannot describe all the flow properties. 
If the morphology of the porous matrix is the same, 
the director pattern is characterized by a non-dimensional 
parameter $W=wL/K$ in the 
absence of flow. 
Two dimensional mapping of the non-equilibrium behaviors 
onto a $W-Er$ plane will 
help to understand the flow properties of NLCs in confined 
systems. 
We need to study them more quantitatively.

We studied the flow behaviors of a NLC 
in a porous medium 
by lattice Boltzmann simulations. 
In the case of strong anchoring, 
the pattern of the director field 
can adopt many (meta-)stable configurations. 
We found that this multistability leads to peculiar 
flow properties of the confined NLCs. 
The dynamic behaviors of topological defects under flow 
depend on whether they are topologically concatenated 
to the matrix.

We also found that 
the apparent viscosity depends on the mutual direction of the flow; 
{\it i.e.,} the NLC flows easily along the 
orientation of the averaged director field. 
Imposing an intermediate flow along an incommensurate direction 
switches the global director field to the new direction. 
Upon this switching, 
the apparent viscosity decreases abruptly to that 
for the commensurate flow. 
Because this defect pattern is maintained even after the flow is stopped, 
we can say the flow path can be recorded in the porous network. 
As reported in a previous paper, 
the emergence of topological defects 
enhance the efficiency of memorization \cite{Araki_NM_2011}. 
We also expect that the 
flow path in a porous matrix can be dynamically 
selected by changing the multistable defect patterns. 
In our simulations, the difference between the apparent viscosities 
of the parallel and perpendicular flows is not large 
compared to that in Ref.~\cite{Care_JPCM_2000}. 
In other words, our system is more isotropic in viscosity. 
This small viscosity difference is attributed to the employed parameters, 
which are restricted in our present scheme. 
When a NLC having a large viscosity anisotropy is used, 
the difference in the apparent viscosity will be more remarkable. 

We employed a bicontinuous matrix with cubic symmetry as 
an example. 
Because the configuration of topological defects depends on the morphology 
of the network, 
our results cannot be directly applied to any matrices. 
However, we consider that dynamic coupling between 
the multiple stabilities and the flow will be found ubiquitously for 
NLCs confined in any porous network. 
We also hope our results will 
improve the understandings of the flow properties of 
NLCs in porous media 
and microfluidic channels.

We acknowledge valuable discussions with H. Tanaka, 
T. Bellini, F. Serra, M. Buscaglia and A. Sengupta. 
This work was supported by the 
JSPS Core-to-Core Program ``International research network 
for non-equilibrium dynamics of soft matter" and KAKENHI. 
The computational work was performed using the facilities at the 
Supercomputer Center, Institute for Solid State Physics, University of Tokyo.

\end{document}